%% file: ms.tex
\documentclass{moriond}

\def\Journal#1#2#3#4{{#1} {\bf #2}, #3 (#4)}

\def\PRL{\em Phys. Rev. Lett.}
\def\PRD{{\em Phys. Rev.} D}

\def\EPJC{{\em Eur. Phys. J.} C}
\def\JHEP{{\em JHEP}}

\def\be{\begin{equation}}
\def\ee{\end{equation}}
\def\bea{\begin{eqnarray}}
\def\eea{\end{eqnarray}}

\usepackage{amsmath}
\usepackage{ifthen} 

\newboolean{uprightparticles}
\setboolean{uprightparticles}{false} 

\input{lhcb-symbols-def}

\newcommand{\Photo}{}

\begin{document}
\vspace*{4cm}
\title{Rare leptonic and semi-leptonic decays at LHCb}

\author{Tom Hadavizadeh, on behalf of the \lhcb collaboration}

\address{Monash University, Melbourne, Australia }

\maketitle\abstracts{
Rare leptonic and semi-leptonic decays offer an excellent opportunity to test the Standard Model of Particle Physics. 
These proceedings report on two recent \lhcb measurements. The first is an observation of the rare $\decay{\jpsi}{\mup\mun\mup\mun}$ decay and the second is a comprehensive measurement of local and nonlocal contributions to the $\decay{\Bz}{\Kstarz\mup\mun}$ decay.
}

\section{Introduction}

Rare decays offer an insightful opportunity to test the Standard Model of Particle Physics (SM). In suppressed decays the impact of physics beyond the SM (BSM) may become apparent, measurably altering the rate or kinematics of the process. 
 Flavour changing neutral currents (FCNC) are a particular set of processes, such as $b \to s \ell \ell $,  which are forbidden at tree-level in the SM, instead proceeding via suppressed topologies such as electroweak penguins. They are an excellent place to search for imprints of new particles that may couple to leptons and quarks. 

Rare decays can provide a wealth of information, including differential decay rates, angular distributions and lepton flavour universality tests. Each measurement differs in the experimental complexity and the precision of theoretical predictions. 
The LHCb experiment is well suited to understanding rare decays of hadrons containing $b$ and $c$ quarks. The precise tracking detectors and efficient hadronic and leptonic particle identification detectors are designed to reconstruct a wide range of processes with a low background contamination. 
These proceedings cover two recent LHCb results~\cite{LHCb-CONF-2024-001,lhcbcollaboration2024comprehensive} shown for the first time at Moriond QCD 2024.

\section{Observation of the rare decay $J/\psi \to \mu^{+}\mu^{-}\mu^{+}\mu^{-}$} 

The decay $J/\psi \to \mu^{+}\mu^{-}\mu^{+}\mu^{-}$ is an electromagnetic process that proceeds through the final-state radiation of a virtual photon.
At present, four lepton decays of heavy quarks are not well studied, however the similarity to FCNC processes, such as $B_s^0 \to \mu^{+}\mu^{-}\mu^{+}\mu^{-}$, make this measurement very useful for understanding final state radiation effects. 
The branching fraction for $J/\psi \to \mu^{+}\mu^{-}\mu^{+}\mu^{-}$ is very precisely predicted in the SM to be $(9.74\pm0.05)\times10^{-7}$.~\cite{PhysRevD.104.094023} 
The \besiii collaboration have observed $J/\psi \to e^{+}e^{-}e^{+}e^{-}$ and $J/\psi \to e^{+}e^{-}\mu^{+}\mu^{-}$ decays and set a limit on $J/\psi \to \mu^{+}\mu^{-}\mu^{+}\mu^{-}$,~\cite{PhysRevD.109.052006} whilst the CMS collaboration has very recently observed $J/\psi \to \mu^{+}\mu^{-}\mu^{+}\mu^{-}$ for the first time, measuring a branching fraction of $(10.1^{3.3}_{2.7}\pm0.4)\times 10^{-7}$.~\cite{cmscollaboration2024observation}

This analysis measures the branching fraction $\mathcal{B}(J/\psi\to\mu^+\mu^-\mu^+\mu^-)$ relative to normalisation channel $J/\psi\to\mu^+\mu^-$ using samples of \jpsi mesons from two different origins.~\cite{LHCb-CONF-2024-001}
A prompt sample includes \jpsi mesons created in the primary proton-proton interaction. However, the high production rate is accompanied by a high background rate, necessitating a tight selection to isolate the signal. A second sample is made up of secondary \jpsi mesons from the decays of \bquark-hadrons. Although this has a lower production rate, the isolated topology allows a significant background reduction, whilst profitting from the existing $B$-decay triggers.  
In each case dedicated BDT algorithms are trained to reject combinatorial background. 

\begin{figure}
\begin{minipage}{1.0\linewidth}
\centerline{\includegraphics[width=0.7\linewidth]{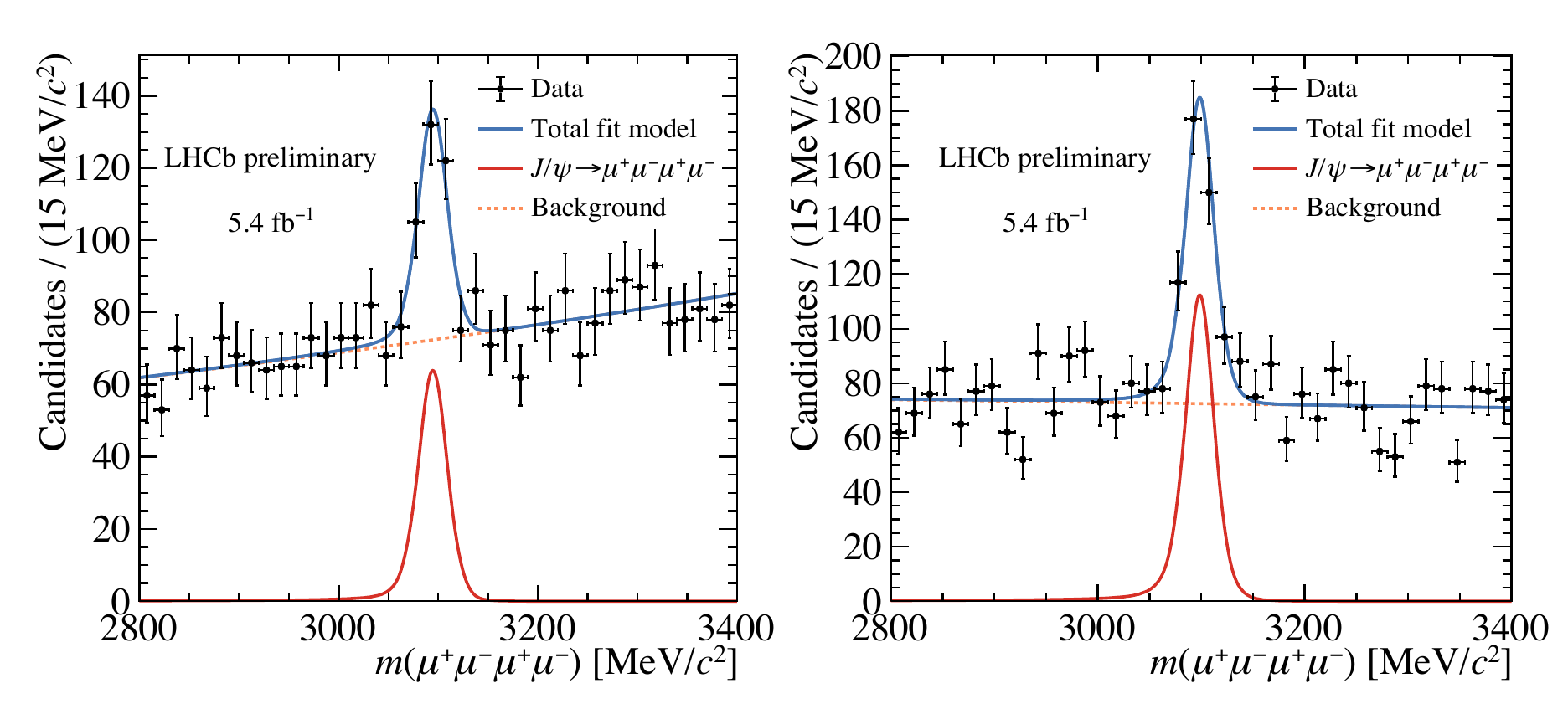}}
\end{minipage}
\hfill
\begin{minipage}{1.0\linewidth}
\centerline{\includegraphics[width=0.7\linewidth]{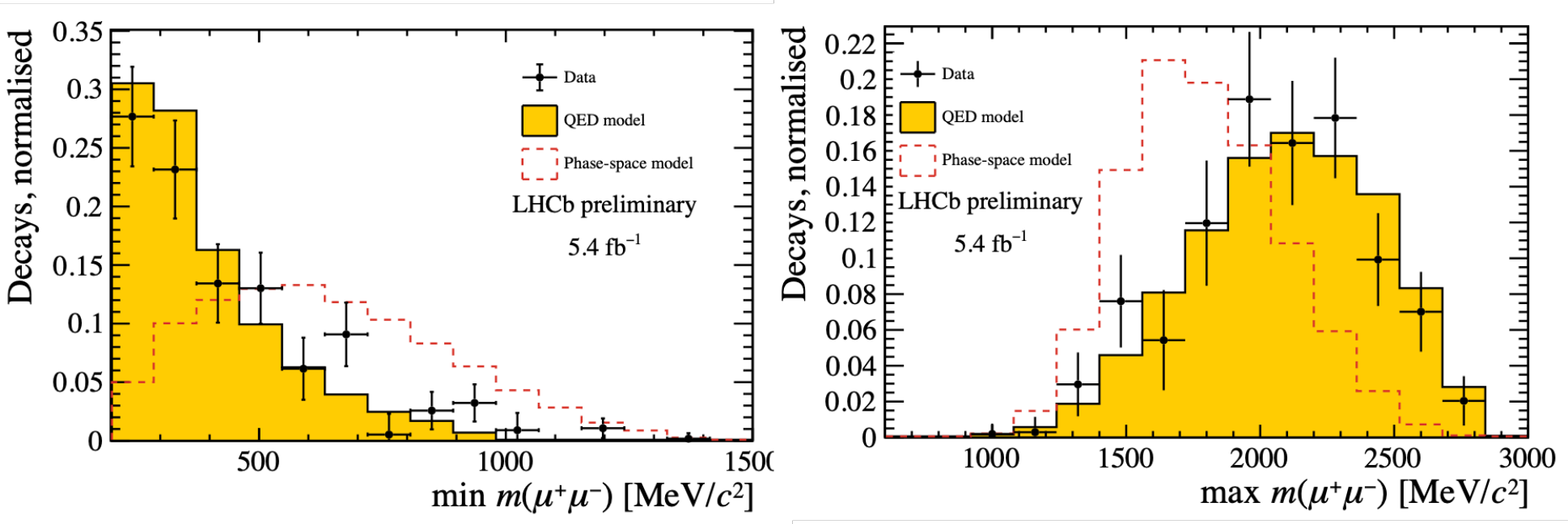}}
\end{minipage}
\caption[]{(Top) Invariant mass distribution of (left) prompt and (right) secondary $J/\psi \to \mu^+ \mu^- \mu^+ \mu^- $ candidates. (Bottom) Dimuon mass distributions compared to a phasespace and QED model~\cite{PhysRevD.109.052006}.  }
\label{fig:jpsi}
\end{figure}

The decay $J/\psi \to \mu^+ \mu^- \mu^+ \mu^- $ is observed in both samples with a large significance ($\gg 5 \sigma $), shown in Fig.~\ref{fig:jpsi}, corresponding to a relative and absolute branching fraction of
\bea
\frac{\mathcal{B}(J/\psi\to \mu^+ \mu^-  \mu^+ \mu^-)}{\mathcal{B}(J/\psi\to \mu^+ \mu^- )}  &= & (1.89\pm 0.17\pm0.09)\times 10^{-5},~ {\rm and} \nonumber\\
\mathcal{B}(J/\psi\to \mu^+ \mu^-  \mu^+ \mu^-)&=& (11.3\pm 1.0\pm0.5\pm0.1)\times10^{-7},\nonumber
\eea
respectively. This is the most precise measurement to date, and consistent with the SM within $1.4\sigma$. The dimuon mass distributions are in agreement with QED predictions as shown in Fig.~\ref{fig:jpsi}.

\section{Comprehensive analysis of local and nonlocal amplitudes in the $B^0 \to K^{*0} \mu^+ \mu^-$ decay}

$B^0 \to K^{*0} \mu^+ \mu^-$ has been subject to much interest over the past 10 years, as the differential decay rate and many angular observables have shown tensions with SM predictions. At the heart of the decay is the quark-level  $b \to s \ell^+ \ell^- $ FCNC transition that can be described in the framework of a weak effective theory. The Wilson coefficients (WC) $\mathcal{C}^{(')}_7$, $\mathcal{C}^{(')}_9$ and $\mathcal{C}^{(')}_{10}$ quantify the left-handed (right-handed) electromagnetic, vector and axial-vector contributions, respectively. The current measurements point to an anomalous vector contribution~\cite{Alguer__2023}.  

The $B^0 \to K^{*0} \mu^+ \mu^-$  final state, however, receives contributions from other, more abundant processes, referred to as charm-loop contributions. Many of the contributions are vector-like, which can modify the measured value of $\mathcal{C}_9$, mimicking the effect of BSM physics. This new analysis models  charm-loop contributions, including both one particle resonances and two particle contributions, allowing an analysis that is unbinned in $q^{2}\equiv m^{2}(\mup\mun)$ across the full  $0.1 <q^2 < 18.0 \gevgevcccc $ range.

Previous \lhcb measurements have investigated $B^0 \to K^{*0} \mu^+ \mu^-$ decays using data corresponding to 4.7\invfb of integrated luminosity, collected during Run 1 (2011-2012) and 2016. These used binned~\cite{LHCb:2020lmf} and unbinned~\cite{LHCb:2023gel} methodologies. This new analysis is the first to used the full Run 1 and Run 2 (2016-2018) dataset and corresponds to 8.4\invfb of integrated luminosity.

The decay rate of the rare $B^0 \to K^{*0} \mu^+ \mu^-$ P-wave process (where $\Kstarz$ refers to the $K^{*0}(892)$) is described using transversity amplitudes that depend on the WCs and $B \to K^{*}$ form factors~\cite{Gubernari_2022}. Full details of the model can be found in Ref.~\cite{lhcbcollaboration2024comprehensive}. 
The nonlocal model comprises all vector resonances that couple to muons, as well as two particle contributions from $D^{(*)}\overline{D}^{(*)}$ and $\tau^+\tau^-$ loops. 
These lead to a $q^2$- and helicity-dependent ($\lambda$) shift in the effective $\mathcal{C}_9$, where the nonlocal terms, $Y$, are calculated using a subtracted hadronic dispersion relation~\cite{Khodjamirian_2013,Cornella_2020}
\bea
\mathcal{C}_9^{\rm{eff},\lambda}(q^2)  = 
\mathcal{C}^\mu_9 + 
Y_{c\bar{c}}^{(0),\lambda} +  
Y_{c\bar{c}}^{1P,\lambda}(q^2)+ 
Y_{\rm{light}}^{1P,\lambda}(q^2)+ 
Y_{c\bar{c}}^{2P,\lambda}(q^2) +
Y_{\tau\bar{\tau}}(q^2).
\eea
The term $Y_{c\bar{c}}^{(0),\lambda}$ represents a subtraction constant whose value is taken from calculations~\cite{Asatrian_2020}. 
The model includes a contribution, $Y_{\tau\bar{\tau}}(q^2)$, from $B^0 \to K^{*0} \tau^+ \tau^-$ decays with $[\tau^+ \tau^- \to \gamma^{*} \to \mu^+ \mu^-]$ rescattering, allowing for the first direct measurement of $\mathcal{C}_{9\tau}$.
A further helicity-dependent shift is introduced in $\mathcal{C}_7^{{\rm eff},\lambda} = \mathcal{C}_7 + \Delta \mathcal{C}_7^\lambda$ to allow for an additional $q^2$-independent helicity-dependent complex phase.

\begin{figure}
    \centering
    \includegraphics[width=0.6\linewidth]{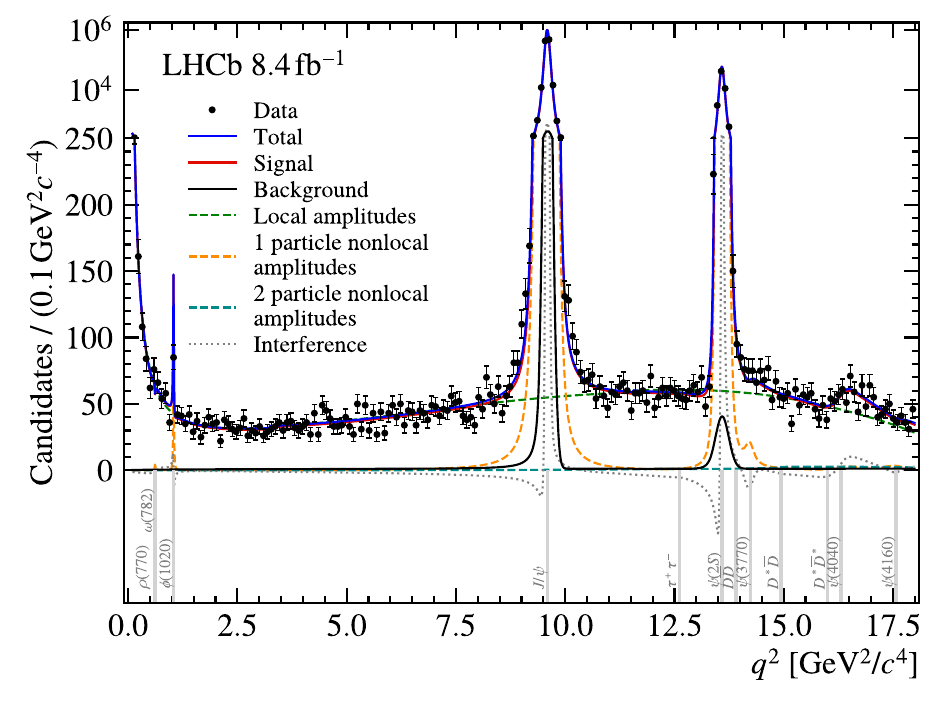}
    \caption[]{Distribution of $q^2$ for $B^0 \to K^{*0} \mu^+ \mu^-$ candidates, shown with different local and nonlocal components of the fitted amplitude model. }
    \label{fig:b2kstarmumu_fit}
\end{figure}

An angular analysis is performed using three decay angles $\Omega = (\cos \theta_K, \cos\theta_\ell, \phi)$ and $q^2$. An acceptance function is determined from simulation to account for experimental efficiencies. Data driven approaches are used to quantify the background distributions and $q^2$ resolution model, both performed in three separate regions of $q^2$. Only data within $0.796 < m_{K\pi} < 0.996 \gevcc$ are used in this analysis. A decoupled model is used to account for S-wave contributions from $ B^0 \to K_0^{*0}(700) \mu^+ \mu^- $ decays. 
The fit model comprises 150 parameters that are allowed to vary freely, including the WCs and nonlocal model parameters. The local $B \to K^{*}$ form factors are allowed to vary within a  Gaussian constraint derived from theoretical uncertainties~\cite{Gubernari_2022}.

\begin{figure}
\centering
\begin{minipage}{0.59\linewidth}
\centerline{\includegraphics[width=1.0\linewidth]{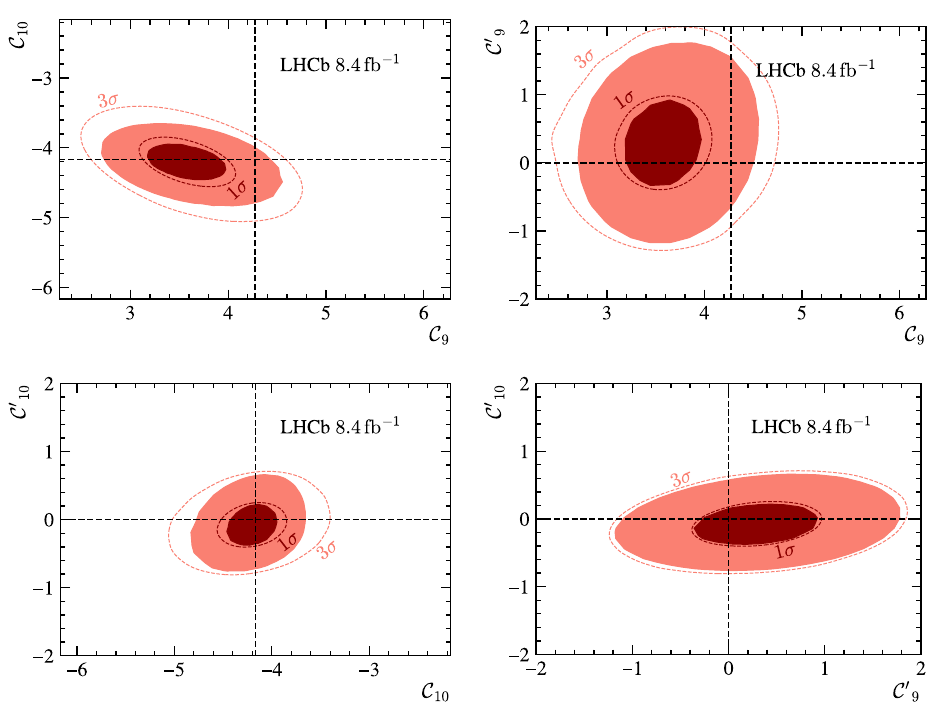}}
\end{minipage}
\begin{minipage}{0.39\linewidth}
\centerline{\includegraphics[width=1.0\linewidth]{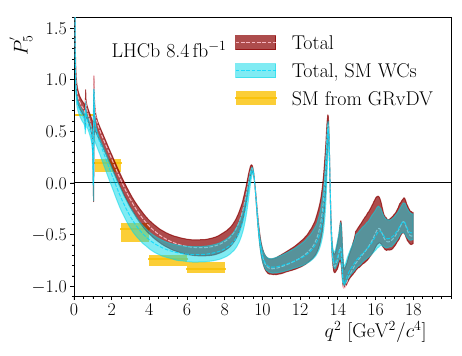}}
\end{minipage}
\caption[]{(Left and Middle) Two dimensional $1\sigma$ and $3\sigma$ contours for $\mathcal{C}^{(')}_9$ and $\mathcal{C}^{(')}_{10}$. The contours are shown representing just statistical uncertainties (shaded) and both statistical and systematic uncertainties (dotted). (Right) The $P'_5$ angular observable as a function of $q^2$ obtained from the fit to data (red), compared to SM model predictions (yellow) and the fit result with the WC set to the SM values (cyan).}
\label{fig:b2kstarmumu_contours}
\end{figure}

The fit to the data, shown in Fig.~\ref{fig:b2kstarmumu_fit}, determines the WC to be 
\bea
\mathcal{C}_9 & = & \phantom{+}3.56\pm 0.28\pm 0.18,  \nonumber \\
\mathcal{C}_{10}&=  & -4.02\pm 0.18\pm0.16,\nonumber \\
\mathcal{C}'_{9} & = & \phantom{+}0.28\pm 0.41\pm 0.12,  \nonumber \\
\mathcal{C}'_{10}&=  & -0.09\pm 0.21\pm0.06,\nonumber \\
\mathcal{C}_{9\tau}&=  & -(1.0\pm 2.6\pm1.0)\times10^2\nonumber 
\label{eq:sp}
\eea
The biggest difference from the SM is observed in $\mathcal{C}_9$ where $\Delta \mathcal{C}_{9}^{\rm BSM} \equiv \mathcal{C}_9 - \mathcal{C}_9^{\rm SM} = -0.71$ which is $2.1\sigma$ from the SM value. The global significance is approximately $1.5\sigma$ with respect to the SM. The results are in agreement with the previous unbinned analysis. 
The systematic uncertainty is dominated by the uncertainty in the branching fraction $\BR(\decay{\Bz}{\jpsi\Kstarz})$ used for normalisation. 

The nonlocal amplitudes are observed to modify the values of the effective Wilson coefficients as a function of $q^2$, in agreement with the previous unbinned analysis~\cite{LHCb:2023gel}. The resulting angular observable $P'_5$ is shown in Fig.~\ref{fig:b2kstarmumu_contours}. The results indicate that the nonlocal contributions, while important, are not sufficient to explain the deviations in the measured value of $\mathcal{C}_9$.
 
\section{Conclusions}
The \lhcb collaboration has performed two new analyses of rare decay processes. The first constitutes the most precise measurement of $\BR(J/\psi \to \mu^{+}\mu^{-}\mu^{+}\mu^{-})$ to date.
The second is a measurement of the local and nonlocal contributions to $B^0 \to K^{*0} \mu^+ \mu^-$ decays, including the first direct measurement of $\mathcal{C}_{9\tau}$. It is found that the nonlocal contributions are not sufficient to explain the deviations observed with the SM predictions.

\end{document}

%% file: lhcb-symbols-def.tex
\usepackage{xspace} 
\usepackage{upgreek}

\def\lhcb   {\mbox{LHCb}\xspace}

\def\besiii {\mbox{BESIII}\xspace}

\def\MagUp {\mbox{\em Mag\kern -0.05em Up}\xspace}

\ifthenelse{\boolean{uprightparticles}}%
{

 \def\Pmu         {\ensuremath{\upmu}\xspace}

 \def\Ppsi        {\ensuremath{\uppsi}\xspace}

 \def\PDelta      {\ensuremath{\Delta}\xspace}                 
 \def\PXi         {\ensuremath{\Xi}\xspace}                 
 \def\PLambda     {\ensuremath{\Lambda}\xspace}                 
 \def\PSigma      {\ensuremath{\Sigma}\xspace}                 
 \def\POmega      {\ensuremath{\Omega}\xspace}                 
 \def\PUpsilon    {\ensuremath{\Upsilon}\xspace}

 \def\PB      {\ensuremath{\mathrm{B}}\xspace}                 
                  
 \def\PD      {\ensuremath{\mathrm{D}}\xspace}

 \def\PJ      {\ensuremath{\mathrm{J}}\xspace}                 
 \def\PK      {\ensuremath{\mathrm{K}}\xspace}

 \def\Pb      {\ensuremath{\mathrm{b}}\xspace}

 \def\Pi      {\ensuremath{\mathrm{i}}\xspace}

 \def\Ps      {\ensuremath{\mathrm{s}}\xspace}

 \def\thebaroffset{0.0em}
}
{

 \def\Pmu         {\ensuremath{\mu}\xspace}

 \def\Ppsi        {\ensuremath{\psi}\xspace}                 
                  
 \mathchardef\PDelta="7101
 \mathchardef\PXi="7104
 \mathchardef\PLambda="7103
 \mathchardef\PSigma="7106
 \mathchardef\POmega="710A
 \mathchardef\PUpsilon="7107
                  
 \def\PB      {\ensuremath{B}\xspace}                 
                  
 \def\PD      {\ensuremath{D}\xspace}

 \def\PJ      {\ensuremath{J}\xspace}                 
 \def\PK      {\ensuremath{K}\xspace}

 \def\Pb      {\ensuremath{b}\xspace}

 \def\Pi      {\ensuremath{i}\xspace}

 \def\Ps      {\ensuremath{s}\xspace}

 \def\thebaroffset{0.18em}
}
\newcommand{\offsetoverline}[2][\thebaroffset]{\kern #1\overline{\kern -#1 #2}}%

\makeatletter
\ifcase \@ptsize \relax
  \newcommand{\miniscule}{\@setfontsize\miniscule{4}{5}}
\or
  \newcommand{\miniscule}{\@setfontsize\miniscule{5}{6}}
\or
  \newcommand{\miniscule}{\@setfontsize\miniscule{5}{6}}
\fi
\makeatother

\DeclareRobustCommand{\optbar}[1]{\shortstack{{\miniscule (\rule[.5ex]{1.25em}{.18mm})}
  \\ [-.7ex] $#1$}}

\def\mup        {{\ensuremath{\Pmu^+}}\xspace}
\def\mun        {{\ensuremath{\Pmu^-}}\xspace} 

\def\squark    {{\ensuremath{\Ps}}\xspace}

\def\bquark    {{\ensuremath{\Pb}}\xspace}

\def\kaon    {{\ensuremath{\PK}}\xspace}

\def\KorKbar {\kern \thebaroffset\optbar{\kern -\thebaroffset \PK}{}\xspace}

\def\Kstarz  {{\ensuremath{\kaon^{*0}}}\xspace}

\def\D       {{\ensuremath{\PD}}\xspace}

\def\DorDbar {\kern \thebaroffset\optbar{\kern -\thebaroffset \PD}\xspace}

\def\Dp      {{\ensuremath{\D^+}}\xspace}
\def\Dm      {{\ensuremath{\D^-}}\xspace}

\def\DpDm    {\ensuremath{\Dp {\kern -0.16em \Dm}}\xspace}

\def\B       {{\ensuremath{\PB}}\xspace}

\def\BorBbar {\kern \thebaroffset\optbar{\kern -\thebaroffset \PB}\xspace}
\def\Bz      {{\ensuremath{\B^0}}\xspace}

\def\Bd      {{\ensuremath{\B^0}}\xspace}

\def\BdorBdbar {\kern \thebaroffset\optbar{\kern -\thebaroffset \Bd}\xspace}

\def\Bs      {{\ensuremath{\B^0_\squark}}\xspace}

\def\BsorBsbar {\kern \thebaroffset\optbar{\kern -\thebaroffset \Bs}\xspace}

\def\jpsi     {{\ensuremath{{\PJ\mskip -3mu/\mskip -2mu\Ppsi}}}\xspace}

\def\Y#1S{\ensuremath{\PUpsilon{(#1S)}}\xspace}

\def\LorLbar     {\kern \thebaroffset\optbar{\kern -\thebaroffset \PLambda}\xspace}

\def\BF         {{\ensuremath{\mathcal{B}}}\xspace}
\def\BR         {\BF}

\newcommand{\decay}[2]{\ensuremath{#1\!\to #2}\xspace} 

\def\to                 {\ensuremath{\rightarrow}\xspace}

\def\AT#1     {\ensuremath{A_{\mathrm{T}}^{#1}}\xspace}           

\def\C#1      {\ensuremath{\mathcal{C}_{#1}}\xspace}                       
\def\Cp#1     {\ensuremath{\mathcal{C}_{#1}^{'}}\xspace}                    
\def\Ceff#1   {\ensuremath{\mathcal{C}_{#1}^{\mathrm{(eff)}}}\xspace}        
\def\Cpeff#1  {\ensuremath{\mathcal{C}_{#1}^{'\mathrm{(eff)}}}\xspace}       
\def\Ope#1    {\ensuremath{\mathcal{O}_{#1}}\xspace}                       
\def\Opep#1   {\ensuremath{\mathcal{O}_{#1}^{'}}\xspace}                    

\newcommand{\aunit}[1]{\ensuremath{\text{\,#1}}}       

\newcommand{\tev}{\aunit{Te\kern -0.1em V}\xspace}
\newcommand{\gev}{\aunit{Ge\kern -0.1em V}\xspace}
\newcommand{\mev}{\aunit{Me\kern -0.1em V}\xspace}
\newcommand{\kev}{\aunit{ke\kern -0.1em V}\xspace}
\newcommand{\ev}{\aunit{e\kern -0.1em V}\xspace}
 
\newcommand{\mevc}{\ensuremath{\aunit{Me\kern -0.1em V\!/}c}\xspace}
\newcommand{\gevc}{\ensuremath{\aunit{Ge\kern -0.1em V\!/}c}\xspace}
\newcommand{\mevcc}{\ensuremath{\aunit{Me\kern -0.1em V\!/}c^2}\xspace}
\newcommand{\gevcc}{\ensuremath{\aunit{Ge\kern -0.1em V\!/}c^2}\xspace}
\newcommand{\gevgevcccc}{\ensuremath{\gev^2\!/c^4}\xspace} 

\def\fb   {\ensuremath{\aunit{fb}}\xspace}
\def\invfb   {\ensuremath{\fb^{-1}}\xspace}

\def\gsim{{~\raise.15em\hbox{$>$}\kern-.85em
          \lower.35em\hbox{$\sim$}~}\xspace}
\def\lsim{{~\raise.15em\hbox{$<$}\kern-.85em
          \lower.35em\hbox{$\sim$}~}\xspace}

\def\tell1  {TELL1\xspace}
\def\ukl1   {UKL1\xspace}

